\documentclass[12pt]{article}
\usepackage{epsf}
\usepackage{graphicx}

\begin{document}
\title
{Hubble diagrams of soft and hard radiation sources in the
graviton background: to an apparent contradiction between
supernova 1a and gamma-ray burst observations}
\author
{Michael A. Ivanov \\
Physics Dept.,\\
Belarus State University of Informatics and Radioelectronics, \\
6 P. Brovka Street,  BY 220027, Minsk, Republic of Belarus.\\
E-mail: ivanovma@gw.bsuir.unibel.by.}

\maketitle

\begin{abstract} In the sea of super-strong interacting gravitons,
non-forehead collisions with gravitons deflect photons, and this
deflection may differ for soft and hard radiations. As a result,
the Hubble diagram would not be a universal function and it will
have a different view for such sources as supernovae in visible
light and gamma-ray bursts. Observations of these two kinds are
compared here with the limit cases of the Hubble diagram.
\end{abstract}
Keywords: galaxies: distances and redshifts —- cosmology:
observations —- cosmology: theory —- cosmology: distance scale --
gamma-ray bursts: general

\section[1]{Introduction }
After the remarkable observations of supernovae 1a dimming
\cite{2,3}, the standard cosmological model has been changed, and
such the new terms as dark energy and an acceleration of the
expansion are now commonly known. Now another cosmological tool -
gamma-ray bursts observations - makes its debut, but there exists
some contradiction with supernova observations \cite{4}: the
Hubble diagrams for these two kinds of sources are not identical.
In the model by the author \cite{500} based on the conjecture
about an existence of the sea of super-strong interacting
gravitons, supernova observational data may be explained without
dark energy. I would like to show here that this apparent
contradiction between two kinds of observations may be resolved in
my model in a very simple manner: soft and hard radiation sources
may have different Hubble diagrams in it, and for an arbitrary set
of sources, the  Hubble diagram is a multivalued function of a
redshift.
\section[2]{Limit cases of the Hubble diagram in the graviton background}
In the standard cosmological model, the luminosity distance
depends on 1) a redshift which conditions a loss of photon
energies and 2) a history of expansion which defines how big is a
surface on which photons fall. In the model by the author
\cite{500} (there is not any expansion in it), the first factor is
the same, but there are the two new factors: 2') the geometrical
distance $r$ is a non-linear function of a redshift $z$ and 3')
non-forehead collisions with gravitons leads to an additional
relaxation of any photonic flux. Namely, the luminosity distance
is $$D_{L}=a^{-1} \ln(1+z)\cdot (1+z)^{(1+b)/2},$$ where $a=H/c,$
$H$ is the Hubble constant and $c$ is the light velocity. The
theoretical value of relaxation factor $b$ has been found in the
assumption that in any case of a non-forehead collision of a
graviton with a photon, the latter leaves a photon flux detected
by a remote observer (the assumption of a narrow beam of rays -
but it is not a well-chosen name): $b=2.137$. It is obvious that
this assumption should be valid for a soft radiation when a photon
deflection angle is big enough and collisions are rare.
\par It is easy to find a value of the factor $b$ in another
marginal case - for a very hard radiation. Due to very small
ratios of graviton to photon momenta, photon deflection angles
will be small, but collisions will be frequent because the
cross-section of interaction is a bilinear function of graviton
and photon energies in this model. It means that in this limit
case $b \rightarrow 0.$
\par For an arbitrary source spectrum, a value of the factor $b$
should be still computed, and it will not be a simple task. It is
clear that $0 \leq b \leq 2.137,$ and in a general case it should
depend on a rest-frame spectrum and on a redshift. It is important
that the Hubble diagram is a multivalued function of a redshift:
for a given $z,$ $b$ may have different values.
\par Theoretical distance moduli $\mu_{0}(z) = 5 \log D_{L} + 25$ are
shown in Fig. 1 for $b=2.137$ (solid), $b=1$ (dot) and $b=0$
(dash). If this model is true, all observations should lie in the
\begin{figure}[th]
\epsfxsize=12.98cm \centerline{\epsfbox{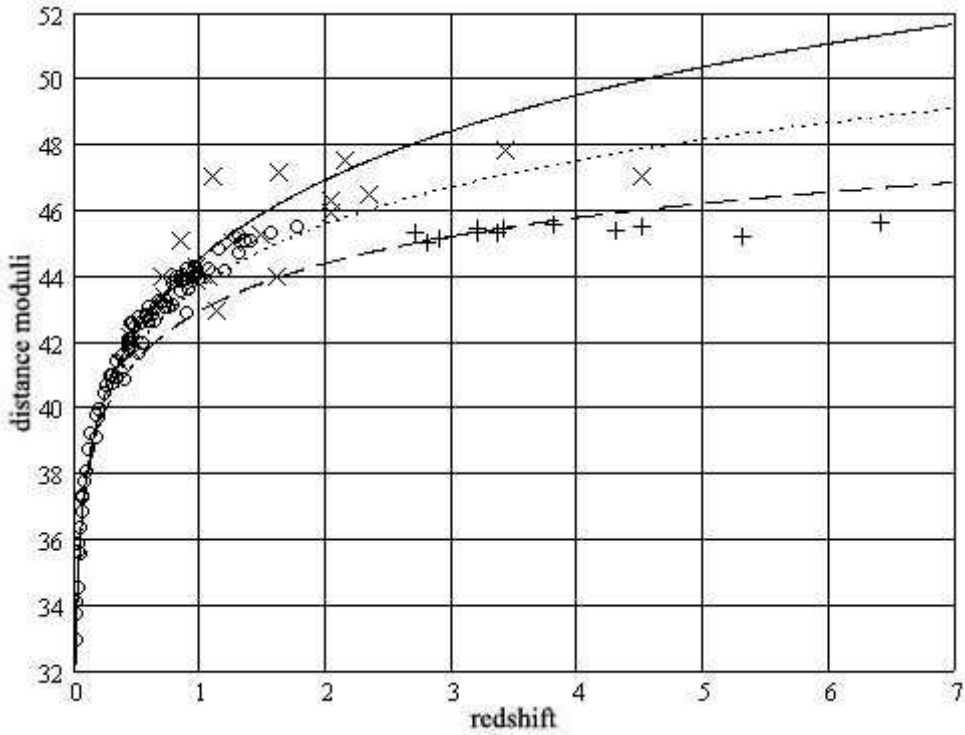}} \caption{
Hubble diagrams $\mu_{0}(z)$ with $b=2.137$ (solid), $b=1$ (dot)
and $b=0$ (dash); supernova observational data (circles, 82
points) are taken from Table 5 of \cite{203}, gamma-ray burst
observations are taken from \cite{204} (x, 24 points) and from
\cite{205} (+, 12 points for $z>2.6$).}
\end{figure}
stripe between lower and upper curves. For Fig. 1, supernova
observational data (circles, 82 points) are taken from Table 5 of
\cite{203}, gamma-ray burst observations are taken from \cite{204}
(x, 24 points) and from \cite{205} (+, 12 points for $z>2.6$). As
it was recently shown by Cuesta et al. \cite{4}, the Hubble
diagram with $b=1$ (in the language of this paper) gives the best
fit to the full sets of gamma-ray burst observations of
\cite{204,205} and it takes place in the standard FLRW cosmology
plus the strong energy condition. Twelve observational points of
\cite{205} belong to the range $z>2.6$, and one can see (Fig. 1)
that these points peak up the curve with $b=0$ which corresponds
in this model to the case of very hard radiation in the
non-expanding Universe with a flat space. In a frame of models
without expansion, any red-shifted source may not be brighter than
it is described with this curve.
\par
\begin{figure}[th]
\epsfxsize=12.98cm \centerline{\epsfbox{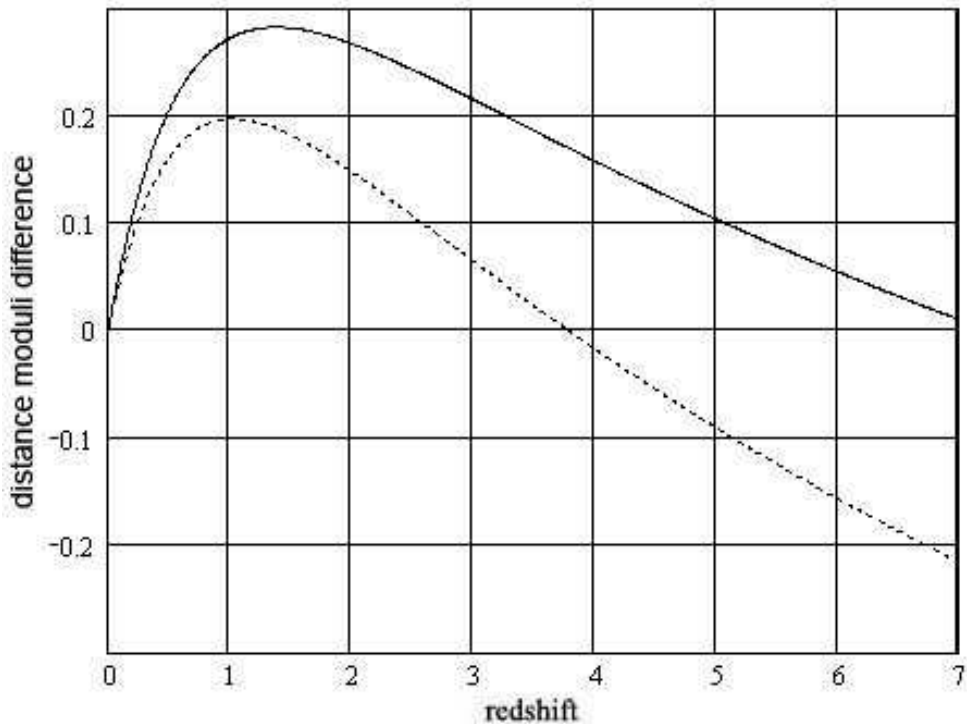}} \caption{ The
difference  $\mu_{c}(z)-\mu_{0}(z)$ for $b=1$ (solid) and $b=1.1$
(dot).}
\end{figure}
Very recently, Schaefer \cite{206} has published a collection of
69 gamma-ray burst observations where calculated distance moduli
are {\it model-dependent}: some cosmological model is used {\it to
calculate the luminosity distance} which is used to evaluate
parameters of bursts. When one compares - after it - GRB
observations with the used cosmological model constructing the
Hubble diagram, one is restricted to be able to check only the
self-consistency of the initial conjecture that the chosen model
is true. As it is shown in Fig. 2, theoretical distance moduli
$\mu_{c}(z)$ for a flat Universe with the concordance cosmology
\begin{figure}[th]
\epsfxsize=12.98cm \centerline{\epsfbox{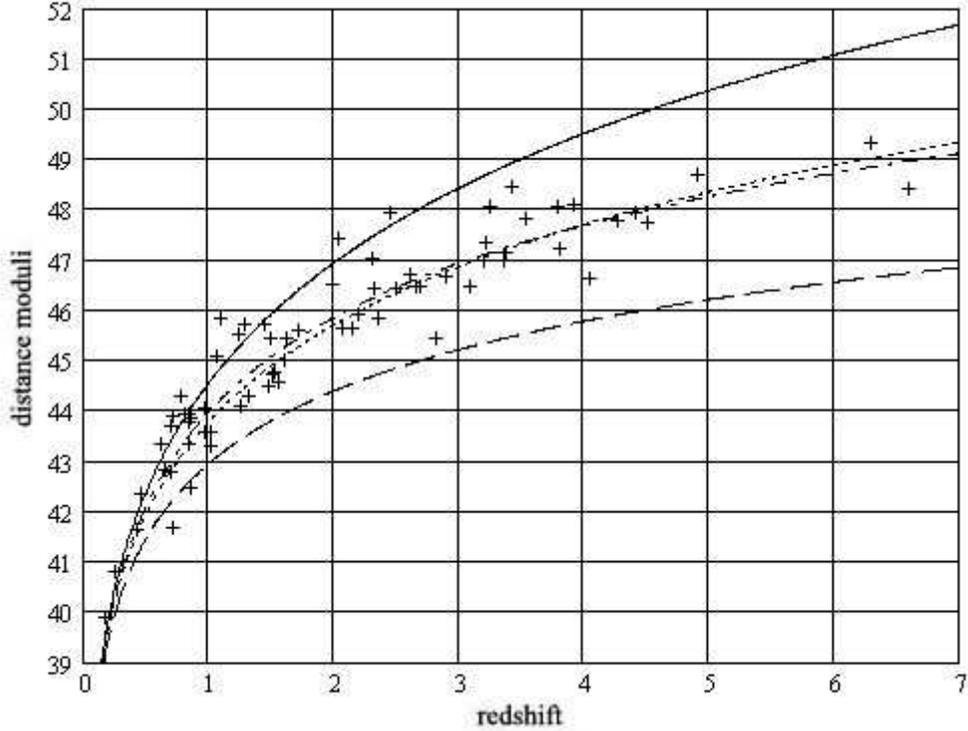}} \caption{ The
same as in Fig. 1 Hubble diagrams $\mu_{0}(z)$ with $b=2.137$
(solid) and $b=0$ (dash); the Hubble diagrams $\mu_{0}(z)$ with
$b=1.1$ of this model (dot) and the one of the concordance model
(dadot) which is the best fit to observations \cite{206}; GRB
observational data (+, 69 points) are taken from Table 6
($\mu^{a}$) of \cite{206} by Schaefer.}
\end{figure}
with $\Omega_{M} = 0.27$ and $w =-1$, which give the best fit to
observations \cite{206}, are very close to the Hubble diagram
$\mu_{0}(z)$ with $b=1.1$ of this model (the difference is not
bigger than $\pm 0.2$ mag in the range $z\leq 6.6$). Because of
this, I would like to compare his calculated GRB distance moduli
for a flat Universe with the concordance cosmology (see Table 6 of
\cite{206}) with theoretical predictions of the considered model
in Fig. 3. We can see that GRB observations lie in the stripe
between lower and upper curves of this model, and the curve
$\mu_{0}(z)$ with $b=1.1$ (or with some bigger $b$) may replace
$\mu_{c}(z)$ with a success. But this curve {\it is not} the limit
case for a very hard radiation. Comparing GRB observational points
on Fig. 1 and Fig. 3 for the same range of $z > 2.6$, we see also
that distance moduli of the last set are essentially higher than
the ones reported in \cite{205} by the same author.
\par
Improved distances to nearby type Ia supernovae (for the range $z
< 0.14$) can be fitted with the function $\mu_{c}(z)$ for a flat
Universe with the concordance cosmology with $\Omega_{M} = 0.30$
and $w =-1$ \cite{207}. In Fig. 4, the difference
$\mu_{c}(z)-\mu_{0}(z)$ between this function and distance moduli
in the considered model is shown for $b=1.52$ (solid), $b=1.51$
(dot) and $b=1.53$ (dash). For $b=1.52$, this difference has the
order of $\pm 0.001$ in the considered range of redshifts.
\begin{figure}[th]
\epsfxsize=12.98cm \centerline{\epsfbox{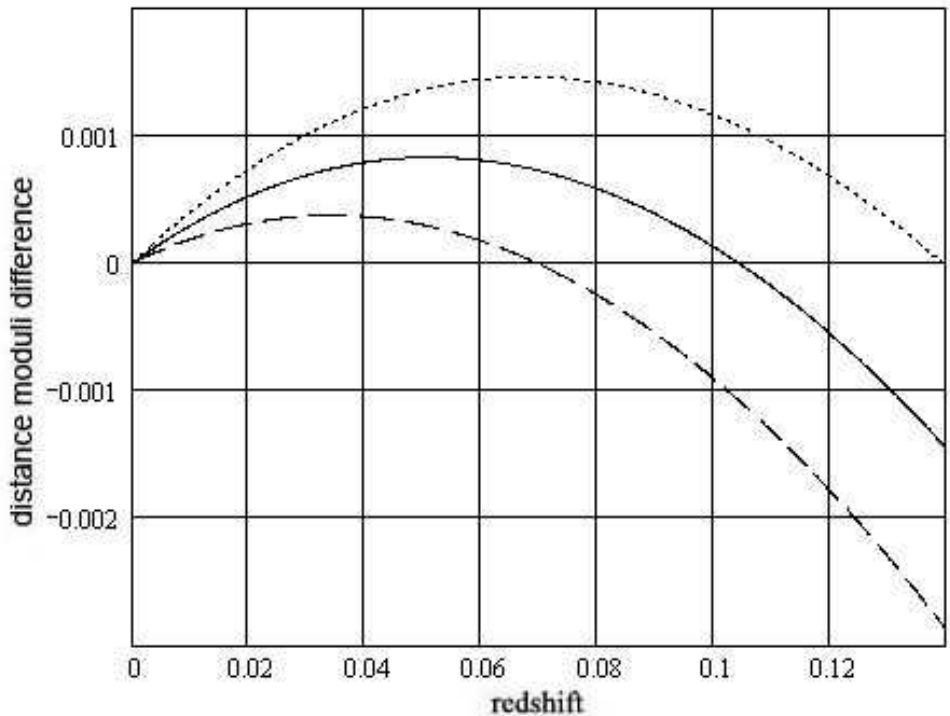}} \caption{ The
difference  $\mu_{c}(z)-\mu_{0}(z)$ for $b=1.52$ (solid), $b=1.51$
(dot) and $b=1.53$ (dash); $\mu_{c}(z)$ corresponds to a flat
Universe with the concordance cosmology with $\Omega_{M} = 0.30$
and $w =-1$, which gives the best fit to supernova observations
for small redshifts \cite{207}.}
\end{figure}
\par
Results from the ESSENCE Supernova Survey together with other
known supernovae 1a observations in the bigger redshift range
$z<1$ can be best fitted in a frame of the concordance cosmology
in which $\Omega_{M} \simeq 0.27$ and $w =-1$ \cite{208}; the
function $\mu_{c}(z)$ for this case is almost indistinguishable
from distance moduli in the considered model for $b=1.405$. In
Fig. 5, the difference $\mu_{c}(z)-\mu_{0}(z)$ is shown for
$b=1.405$ (solid), $b=1.400$ (dot) and $b=1.410$ (dash). For
$b=1.405$, this difference is not bigger than $\pm 0.035$ for
redshifts $z<1$ (the same is true for slightly different values of
$\Omega_{M}$ used in \cite{208}, too,  but for some other values
of the factor $b$: for $\Omega_{M}=0.274$ or 0.267,  $b$ is equal
to 1.400 or 1.410 correspondingly).
\begin{figure}[th]
\epsfxsize=12.98cm \centerline{\epsfbox{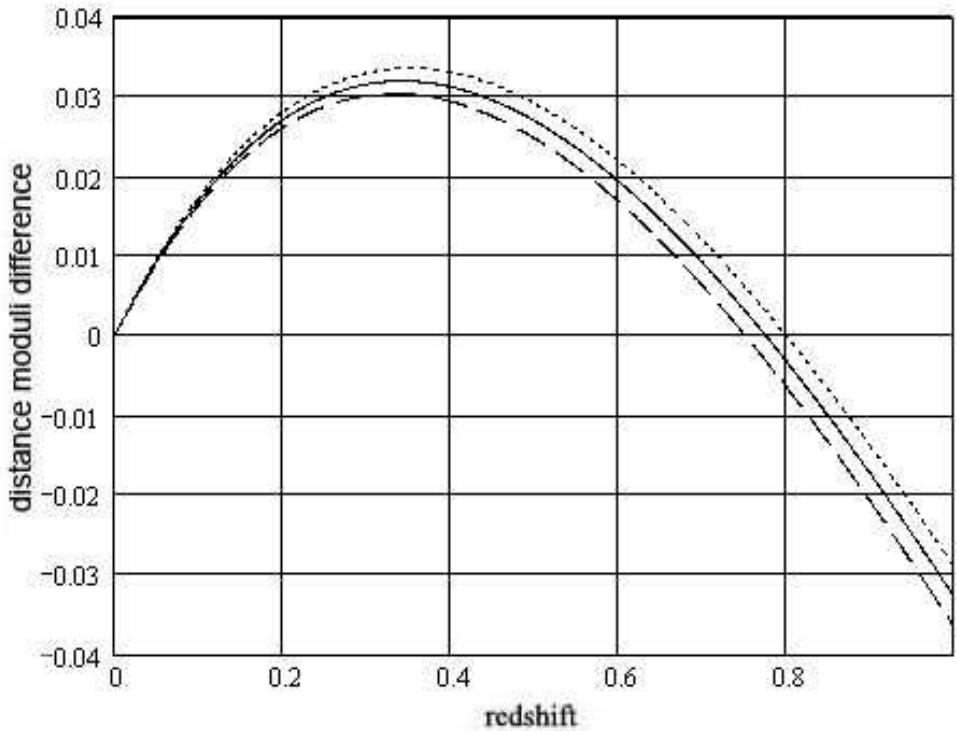}} \caption{ The
difference  $\mu_{c}(z)-\mu_{0}(z)$ for $b=1.405$ (solid),
$b=1.400$ (dot) and $b=1.410$ (dash); $\mu_{c}(z)$ corresponds to
a flat Universe with the concordance cosmology with $\Omega_{M} =
0.27$ and $w =-1$, which gives the best fit to supernova
observations for the bigger redshift range $z<1$ \cite{208}.}
\end{figure}
\par The gold sample of supernovae \cite{203} by Riess et al. has
the best fit with $w(z)=w_{0}+w^{'}z,$ where $ w_{0}=-1.31$ and
$w^{'} = 1.48$ (dark energy changes with redshift); because this
supernovae Hubble diagram goes below of the GRB one \cite{206} for
$z>1$, and in a frame of the considered model it is {\it
impossible}, it may be that the GRB derived distance moduli by
Schaefer \cite{206} are not consistent now with the supernovae
observations.
\section[3]{Conclusion}
The considered multivalued character of the Hubble diagram may
explain an apparent contradiction between supernovae and GRBs
observations. We have now a very poor set of GRBs with big
redshifts, and it is obvious that errors of observations are very
large. When such missions as the SWIFT satellite observe much more
GRBs at high redshifts, one can get a surprising result:
observations would lie on the curve which corresponds to the
non-expanding Universe. It would be very important to get
supernova data for higher redhifts with the help of new missions
to be able to do more definitive conclusions.

\end{document}